\documentclass[prc,onecolumn,tightenlines,12pt]{revtex4}
\usepackage{graphicx}
\usepackage{dcolumn}
\usepackage{amsmath}
\newcommand{\be}{\begin{equation}}
\newcommand{\ee}{\end{equation}}
\newcommand{\bea}{\begin{eqnarray}}
\newcommand{\eea}{\end{eqnarray}}
\newcommand{\f}{\frac}
\newcommand{\la}{\lambda}
\newcommand{\La}{\Lambda}

\begin{document}

\author{M. S. Hussein$^{\dag 1,2}$, 
J. X. de Carvalho$^{1,2}$,  
M. P. Pato${^2}$ and A. J. Sargeant$^{2}$}
\affiliation{$^{1}$Max-Planck-Institut f\"ur Physik komplexer Systeme\\
N\"othnitzer Stra$\beta$e 38, D-01187 Dresden, Germany \\
$^{2}$Instituto de F\'{i}sica, Universidade de S\~{a}o Paulo\\
C.P. 66318, 05315-970 S\~{a}o Paulo, S.P., Brazil}
\title{Symmetry Breaking Study with  Random Matrix Ensembles
\thanks{Supported in part by the CNPq and FAPESP (Brazil).\\
        $^{\dag}$Martin Gutzwiller  Fellow, 2007/2008.}}

\begin{abstract}

A random matrix model to describe the coupling of $m$-fold symmetry 
is constructed. The particular threefold case is used to analyze 
data on eigenfrequencies of elastomechanical vibration of an 
anisotropic quartz block. It is suggested that such 
experimental/theoretical study may supply a powerful means to
discern intrinsic symmetry of physical systems.
\end{abstract}

\maketitle

The standard ensembles of Random Matrix Theory
(RMT) \cite{Meht} have had wide application
in the description of the statistical properties of eigenvalues and
eigenfunctions of complex many-body systems. Other ensembles have
also been introduced \cite{Dyson}, in order to cover situations that depart
from universality classes of RMT. One such class of ensembles is the
so-called Deformed Gaussian Orthogonal Ensemble (DGOE)
\cite{Pato1,Pato2,Pato3, Carneiro:1991} that proved to be particularly useful when one
wants to study the breaking of a discrete symmetry in a
many-body system such as the atomic nucleus.

In fact, the use of spectral statistics as a probe of symmetries in physical
systems has been a subject of intensive experimental and theoretical
investigation following the pioneering work of Bohigas, Giannoni and Schmit 
\cite{Bohigas}
which showed that the quantal behaviour of classically chaotic systems exhibits the
predictions supplied by the RMT. Examples of symmetry breaking in physical
systems that have been studied include nuclei \cite{Mitch0, Mitch}, atoms 
\cite{Simons, Welch}
and mesoscopic devices such as quantum dots \cite{Alhassid}.

In the case of nuclei, the Mitchell group at the Triangle Universities Nuclear
Laboratory \cite{Mitch0, Mitch}, studied the effect of isospin symmetry breaking, 
in  odd-odd
nuclei such as $^{26}Al$. They detected the breakdown of this important symmetry by
the applications of two statistics: the short-range, nearest neighbor level
spacing distribution (NND) and the long range Dyson's $\Delta$-statistics 
\cite{Mitch0, Mitch}. These results were well described by a DGOE in which a pair of 
diagonal blocks is coupled. The strength of the coupling needed to account for the 
symmetry breaking can be traced to the average matrix element of the Coulomb interaction
responsible for this discrete symmetry breaking \cite{Pato2, Guhr}.  
The justification for the use of block matrices to describe the statistics of
a superposition of $R$ spectra with different values of the conserved
quatum number can be traced to Refs. \cite{Meht, NovaR}.
In the case of non-interacting
spectra, \emph{i.e.} if the quantum number is exactly conserved, the answer is
a superposition of the $R$ spectra. Since the level repulsion is present in
each one of the $R$ spectra, their superposition does not show this feature. 
Thus, we can say that for each spectra of states of a given value 
of the quantum number, one attaches a random matrix (GOE). For $R$ spectra
each of which has a given value of the conserved quantum number, one
would have an $R\times R$ block diagonal matrix. Each block matrix will 
have a dimension dictated by the number state of that spectra. If the
quantum number is not conserved then the $R\times R$ block matrix acquires non-diagonal
matrices that measure the degree of the breaking of the associated symmetry.
This idea was employed by Guhr and Weidenm\"uller \cite{Guhr} and Hussein
and Pato \cite{Pato1} to discuss isopin violation in the nucleus $^{26}$\emph{Al}.
In reference \cite{Pato1}, the random block matrix model was called
the Deformed Gaussian Orthogonal Ensemble (DGOE).

In order to study transitions amongst universal classes of ensembles
such as order-chaos (Poisson$\rightarrow$GOE), symmetry violation 
transitions (2GOE$\rightarrow$1GOE), experiments on physical 
systems are more complicated due to the difficulty of tuning the interaction
(except, e.g. in highly excited atoms where the application of a
magnetic field allows the study of GOE-GUE transitions). To simulate
the microscopic physical systems, one relies on analog computers such
as microwave cavities, pioneered by A. Richter and collaborators
\cite{achim1} and acoustic resonators of Ellegaard and 
collaborators\cite{Elleg0,Elleg,Elleg2}. It is worth mentioning at this
point that the first to draw attention to the applicability of RMT to
accoustic waves in physical system was Weaver \cite{Weaver}.

In the experiment of Ellegaard \emph{et al.} \cite{Elleg} what 
was measured were eigenfrequencies of the elastomechanical vibrations 
of an anisotropic crystal block with a D3 point-group symmetry.
The rectangular crystal block employed by Ellegard was so prepared as
to have only a two-fold flip symmetry retained. Then, to all
effects, the quartz specimen resembles a system of two three-dimensional Sinai
billiards. The statistical treatment of the eigenfrequencies of such a
block would follow that of the superposition of two uncoupled GOE's.

Then, 
by removing octants of  progressively larger radius from a corner of 
the crystal block this remnant two-fold symmetry was gradually broken. 
The spectral statistics show a transition towards 
fully a chaotic system as the octant radius increases. 
What was then seen was that the measured
NND is compatible with a two block DGOE description but the 
$\Delta$-statistics was discrepant. This discrepancy was attributed to 
pseudo integrable behavior and this explanation was later implemented
with the result that the long-range behavior was fitted at the cost, 
however, of loosing the previous agreement shown by the NND\cite{Abul}. 

Here we reanalyse this experiment following the simpler idea of
extending the DGOE matrix model \cite{Pato3} to consider the 
coupling of three instead of two GOE's \cite{Carneiro:1991}. 
We show that 
within this extension both, the short- and the long-range statistics,
are reasonably fitted suggesting that the assumption 
of the reduction of the complex symmetries of anisotropic quartz block
may not be correct. Our findings have the potential of supplying  very
precise means of testing details of symmetry breaking in pysical systems.

To define the ensembles of random matrices we are going to work with, we
recall the construction based on the Maximum Entropy Principle \cite{Pato1},
that leads to a random Hamiltonian which can be cast into the form
\begin{equation}
H=H_{0}+\lambda H_{1},  \label{eq 1b}
\end{equation}
where the block diagonal $H_{0}$ is a matrix made of $m$ uncoupled 
GOE blocks and 
$\lambda $ ($0\leq \lambda \leq 1)$ is the parameter that controls the
coupling among the blocks represented by the $H_{1}$ off-diagonal
blocks. For $\lambda=1,$ the $H_{1}$ part completes the  $H_{0}$ part 
and $H=H^{GOE}.$

These two matrices $H_{0}$ and $H_{1}$ are better expressed introducing the 
following $m$ projection operators 
\begin{equation}
P_{i}=\sum\limits_{j\in I_{i}} \mid j><j\mid,  \label{eq 2a}
\end{equation}
where $I_{i}$ defines the domain of variation of the row and column 
indexes associated with $i$th diagonal block of size $M_i.$
Since we are specifically interested in the transition from 
a set of $m$ uncoupled GOE's to a single GOE,
we use the above projectors to 
generalize our previous model \cite{Pato1,Pato2} by writing
\begin{equation}
H_{0}=\sum\limits_{i=1}^{m}P_{i}H^{GOE}P_{i}  \label{eq 4}
\end{equation}
and
\begin{equation}
H_{1}=\sum\limits_{i=1}^{m}P_{i}H^{GOE}Q_{i}  \label{eq 5}
\end{equation}
where $Q_{i}=1-P_{i}.$ It is easily verified that $H=H^{GOE}$
for $\lambda =1.$

The joint probability distribution of matrix elements can be put 
in the form \cite{Pato1,Alberto}
\begin{equation}
P(H,\alpha,\beta)=Z_{N}^{-1} \exp\left(-\alpha tr H^2 -
\beta tr H_{1}^2\right)  \label{eq 12a}
\end{equation} 
with the parameter $\lambda$ being given in terms of $\alpha$ 
and $\beta$ by

\begin{equation}
\lambda=(1+\beta/\alpha)^{-1/2}.  \label{eq 12b}
\end{equation}

Statistical measures of the completely uncoupled $m$ blocks have been 
derived. They show that level repulsion disappears which can be 
understood since eigenvalues from different blocks behave 
independently. In fact, as $m$ increases the Poisson statistics are
gradually approached. In the interpolating situation of partial 
coupling, some approximate analytical results have been derived.
In Ref. \cite{Alberto}, for instance, it has been found that the 
density $\rho (E)$ for arbitrary $\lambda$ and $m$ is given by
\begin{equation}
\rho(E)=\sum\limits_{i=1}^{m}\frac{M_i}{N}\rho_{i}(E)  \label{eq 5d}
\end{equation}
where 
\begin{equation}
\rho_{i}(E)=\left\{ 
\begin{array}{rl} 
\frac{2}{\pi a_{i}^{2}}\sqrt{{a_{i}^{2}-E^2}}, & \mid E\mid \leq a \\ 
0, &  \mid E \mid   > a
\end{array}
\right. 
\end{equation} \label{eq 5g}
is Wigner's semi-circle law with $a=\sqrt{N/\alpha}$ and
\begin{equation}
  a_{i}=a^{2}\left[\frac{M_{i}}{N}+\lambda^{2}\left(1-\frac{M_{i}}{N}\right)\right].
\end{equation}
The transition parameter utilized in the following is defined as 
\cite{Pandey:1995}
\be
\La=\la^2\rho(0)^2=\la^2\f{N^2}{\pi a/2}.
\ee

Eq.~(\ref{eq 12a}) can be used to calculate exactly analytically 
the NND for $2\times 2$ and 
$3\times 3$ matrices \cite{Carneiro:1991}. For the $2\times 2$ case the
DGOE, Eq. (\ref{eq 12a}), gives 
\begin{equation}
  P_{2\times 2}(s,\Lambda)=s\sqrt{\frac{\pi}{8\Lambda}}I_{0}(\frac{s^{2}}{16\Lambda})
                    \exp\left(-\frac{s^{2}}{16\Lambda}\right),\label{rio}
\end{equation}
where $I_{0}$ is the modified Bessel function, whose asymptotic form is
\begin{equation}
     I_{0}(x) \rightarrow \frac{e^{x}}{\sqrt{2\pi x}}.
\end{equation}
Thus, $P_{2\times 2}(s,0)=1$, and
there is no level repulsion for $\Lambda \rightarrow 0$. 
In the opposite limit, $\Lambda \rightarrow \infty$, 
$I_{0}(x)\approx 1 - x^{2}/4$ and one obtains 
\begin{equation}
  P_{2\times 2}(s, \Lambda \gg 1 )\approx s \sqrt{\frac{\pi}{8\Lambda}}
                         e^{-s^{2}/16\Lambda}. \label{mocambo}
\end{equation}
For higher dimensions Eq.~(\ref{eq 12a}) can only be used for 
numerical simulations, however,
using appropriate perturbative methods Leitner \cite{Leitner} was
able to find a formula for 
the NND. He started basically
with the formula for the nearest neigbhour spacing distribution
for the superposition of $m$ GOE's block matrices \cite{Meht}
\begin{equation}
  P_{m}(s) = \frac{d^{2}}{ds^{2}}E_{m}(s)
\end{equation}
where, for the case of all block marices having the same dimension one
has
\begin{equation}
  E_{m}(s) = \left( E_{1}(\frac{s}{m} ) \right)^{m},
\end{equation}
\begin{equation}
  E_{1}(x) = \int_{x}^{\infty}(1-F(t))\, dt,
\end{equation}
\begin{equation}
  F(t) = \int_{0}^{t}P_{1}(z)\,dz.
\end{equation}
In the above $P_{1}(z)$ is the normalized nearest neighbour spacing
distribution of one block matrix. It is easy to find for $P_{m}(s)$,
the following
\begin{eqnarray}
  P_{m}(s)& =& \frac{1}{m}\left[     \left(E_{1}(s/m)\right)^{m-1}
                                   P_{1}(s/m) + 
                             (m-1)(E_{1}(s/m))^{m-2}(1-F(s/m))^{2}
                        \right] \label{lago1} \\
          & \equiv & P^{(1)}_{m}(s) + P_{m}^{(2)}(s) \label{lago2}
\end{eqnarray}
If all the block matrices belong to the GOE, then one can use the
Wigner form for $P_{1}(z)$
\begin{equation}
  P_{1}(z) = \frac{\pi}{2}z e^{-\frac{\pi}{4}z^{2}}\approx \frac{\pi}{2}z,
\label{noiva1}
\end{equation}
thus
\begin{equation}
  F_{1}(z) = 1 - e^{-\frac{\pi}{4}z^{2}}\approx \frac{\pi}{4}z^{2},
\label{noiva2}
\end{equation}
\begin{equation}
  E_{1}(z) = erfc\left(\frac{\sqrt{\pi}}{2}z \right)\approx 1 - z.
\label{noiva3}
\end{equation}
where the large-$z$ limits of Eqs. (\ref{noiva1})-(\ref{noiva3}) are also
indicated above.
It is now clear that the above expression for $P_{m}(s)$, (\ref{lago1}) and
(\ref{lago2}), contains a term $P_{m}^{(1)}(s)$ with level repulsion,
indicating short-range correlation among levels pertaining to the
same block matrix and a second term $P_{m}^{(2)}(s)$ with no level
repulsion, implying short-range correlation among NND levels
pertaining to different blocks. Notice that for very small spacing, 
$P_{m}(s)$ behaves as 
\begin{equation}
  P_{m}(s) \approx \frac{\pi}{2m^{2}}s + \frac{m-1}{m}
\end{equation}
for $m=1$, we get the usual $P_{1}(0)=0$, while for $m>1$,
we get $P_{m}(0)=(m-1)/m$. To account for symmetry breaking,
Leitner \cite{Leitner} considered the mixing between levels pertaining
to nearest neigbhour block matrices. This amount to constrain
the mixing to be of the form given by Eq. (\ref{rio}) for
the $2\times 2$ DGOE, thus, he found
\begin{equation}
  P_{m}(s,\Lambda)= P_{m}^{(1)}(s)+
                   P_{2\times 2}(s,\Lambda)P_{m}^{(2)}(s). \label{azuis}
\end{equation}
Though $P_{m}(s)$ is normalized, $P_{m}(s,\Lambda)$ is not. Accordingly
one supplies coefficients $c_{N}$ and $c_{D}$, such that
\begin{equation}
  P_{m}(s,\Lambda,c_{N},c_{D})\equiv c_{N}P_{m}(c_{D}s,\Lambda)\label{verde}
\end{equation}
is normalized to unity. Similarly, $<s>$ should be unity too.
Eq. (\ref{azuis}) can certainly be generalized to consider the effect
of mixing of levels pertaining to next to nearest neighbour blocks,
and accordingly, $P_{3\times 3}(s,\Lambda)$, given in Ref. \cite{Carneiro:1991}
would be used
in Eq. (\ref{azuis}) instead of $P_{2\times 2}(s,\Lambda)$. In the 
following, however, we use Eqs. (\ref{azuis}), (\ref{verde}) as Leitner did \cite{Leitner}.

	In Ref. \cite{Leitner}, Leitner also obtained approximate expression 
for the spectral rigidity $\Delta_{3}(L)$ using results derived by 
French \emph{et al.}
\cite{French}. Leitner's approximation to $\Delta_{3}$ is equal to the GOE spectral 
rigidity plus perturbative terms, that is
\begin{eqnarray}
  \Delta_{3}^{(m)}(L;\Lambda)& \approx  & \Delta_{3}(L;\infty) +
        \frac{m-1}{\pi^{2}}\left[
	  \left(\frac{1}{2}-\frac{2}{\epsilon^{2}L^{2}}
                           -\frac{1}{2\epsilon^{4}L^{4}}\right) 
                           \right. \nonumber \\
        & & \times \ln (1+\epsilon^{2}L^{2}) +
            \frac{4}{\epsilon L}\tan^{-1}(\epsilon L ) +
           \left. \frac{1}{2\epsilon^{2}L^{2}} - \frac{9}{4} \right],
                                  \label{asa} 
\end{eqnarray}
where
\begin{eqnarray}
   \epsilon & = & \frac{\pi}{2(\tau + \pi^{2}\Lambda)}
\end{eqnarray}
For the cut off parameter we use the value \cite{Abul-Magd:2004}
$
\tau=c_me^{\pi/8-\gamma-1},
$
where $c_m=m^{m/(m-1)}$ 
and $\gamma\approx 0.5772$ is  Euler's constant.
This choice guarantees that when the symmetry is not
broken, $\Lambda = 0$, $\Delta_{3}^{(m)}(L,0)=m\Delta_{3}(L/m,\infty)$.
In Ref. \cite{Leitner:1997}, Leitner fitted 
Eq. (\ref{verde}) for $m$=2 to the NND from
Ref. \cite{Elleg}, however, he did not fit the spectral rigidity.
It is often the case that there are some missing levels in
the statistical sample analysed. Such a situation was
addressed recently by Bohigas and Pato \cite{ML} who have
shown that if $g$ fraction of the levels or eigenfrequencies
is missing, the $\Delta_{3}(L)$ becomes
\begin{equation}
  \Delta_{3}^{g}(L) = g\frac{L}{15}+(1-g)^{2}\Delta_{3}
                                \left(\frac{L}{1-g}\right).\label{dragao}
\end{equation}
The presence of the linear term, even if small, could explain
the large $L$ behavior of the \emph{measured} $\Delta_{3}(L)$.
We call this effect the Missing Level (ML) effect. Another
possible deviation of $\Delta_{3}$ from Eq. (\ref{asa}) could arise
from the presence of pseudo-integrable effect (PI) \cite{Abul,bis2}.
This also modifies $\Delta_{3}$ by adding a Poisson term 
just like Eq. (\ref{dragao}).

We now apply our model to analyse the eigenfrequency 
data of the elastomechanical vibrations of an anisotropic quartz 
block used in \cite{Elleg}. In this reference in order to break
the flip symmetry of the crystal block gradually they removed
an octant of a sphere of varying size at one of the corners.
The rectangular quartz block has the dimensions 
$14\times 25\times 4\, mm^{3}$. The radii of the spheres containing
the octants are $r=0.0, 0.5, 0.8, 1.1, 1.4$ and $1.7\, mm$ representing
figures $(a)-(f)$. Figs. $1x$ and $2x$ of Ref. \cite{Elleg} 
correspond to the octant $r\gg 1.7$.
They found  1424, 1414, 1424, 1414, 1424 and 1419 frequency eigenmodes,
respectively. The histograms and circles
in the two figures of Ref. \cite{Elleg} represent the short-range
nearest-neighbor distributions $P(s)$ (Fig. 1) and the long range
$\Delta_{3}(L)$ statistics (Fig. 2).

The results of our analysis
are shown in the two figures. In Fig. 1, the 
sequence of six measured NNDs were fitted  for $m=2$ and $m=3.$ 
It can be seen that the DGOE model with three coupled GOE's give a comparable 
and in some cases even better fit than the $m=2$ one. Figure 1a in
fact shows a rather sharp peak in our calculated $P(s)$ for $m=3$,
$P_{3}(s,0.0056)$. We consider this a failure of our formula 
(\ref{verde})
for the uncut crystal. In fact, a more appropriate description of
the uncut crystal is to take $\Lambda = 0$, namely a superposition
of 3 uncoupled GOE's, which works almost as good as the 2 uncoupled GOE's
description. The other parts of figure 1, $(b)-(x)$ seem to show the
same insensitivity of $P_{m}(s,\Lambda)$ to $m$; the number of matrix blocks
used in DGOE description. It is this insensitivity of the short-range
nearest neighbour
level correlation, measured by the spacing distribution, to the assumed
symmetry inherent in the uncut crystal (and thus the number uncoupled
GOE's employed to describe it) that forces us to examine the long-range
level correlation, namely spectral rigidity, ``measured'' by Dyson's
$\Delta_{3}$ statistics.  

In Fig. 2. the 
$\Delta$-statistic was fitted with equation (\ref{asa}).
It is clear from the figure that a good fit to the data
of Ref. \cite{Elleg} is obtained with $m=3$ for the values
of $\Lambda$ given in table 1. This is to be contrasted with
the case of $m=2$  which, according to Eq. (\ref{asa}) results
in $\Delta_{3}^{(2)}(L,\Lambda )$ that is \emph{always} below
the one with $\Lambda=0$, $\Delta_{3}^{(2)}(L,0)$, which itself
is always below the data points of Ref. \cite{Elleg}. For this
reason, only the $\Delta_{3}^{(2)}(L,0)$ is shown in the figure.
It should be noted that the $\Delta$-statistics of the uncut crystal,
Fig. 2a is very well described by that of 3 uncoupled GOE's, 
namely $\Delta_{3}^{(3)}(L)=3\Delta_{3}^{(1)}(L/3)$ which is  
always larger than the above mentioned 
$\Delta_{3}^{(2)}(L)=2\Delta_{3}^{(1)}(L/2)$.
The most conspicuous exception is
 Fig. \ref{presente}b which corresponds to $r=0.5\, mm$
and  where $1414$ frequency eigenvalues were found. We consider
this a potential ML  case and take for $\Delta_{3}$, the
expression given in Eq. (\ref{dragao}) and use it in Eq.
(\ref{asa}). We find perfect fit to the \emph{data}, if $g$
is taken to be $6\%$, namely only $94\%$ of the eigenfrequencies
were in fact taken into account in the statistical analysis. In contrast, if 2GOE is used
we still do not get very good agreement even if 18\% of the levels are taken to be missing, as shown in Fig \ref{presente3}.
There is, threfore, room to account much better for all cases
(Fig. $2a$, $2c$, $\ldots$ ) by appropiately choosing the 
correponding value of $g$.

\begin{table}
\caption{\label{tab:inputs}Values of $\Lambda$ obtained by fitting
    Eqs. (\ref{verde}) and (\ref{asa}) respectively to the
    experimental NNDs and spectral rigidities from Ref. \cite{Elleg}.}
\begin{tabular}{|l|l|l|l|l|}
\hline
&\multicolumn{3}{c|}{$P(s)$}& $\Delta_3(L)$\\
\hline
 Data Set&  Ref. \cite{Leitner:1997}& Eq. (\ref{verde}) $m$=2 & Eq. (\ref{verde}) $m$=3 & Eq. (\ref{asa}) 
$m$=3\\
\hline
(a) & 0.0013 & 0.0030  & 0.0067  & 0.0056 \\
(b) & 0.0054 & 0.0063  & 0.0098  & 0.0016 \\
(c) & 0.0096 & 0.010  & 0.017    & 0.0017 \\
(d) & 0.0313 & 0.032 & 0.064     &  0.027 \\
(e) & 0.0720 & 0.070 & 0.13 & 0.050 \\
(f) & 0.113 & 0.12& 0.30 & 0.16 \\
(x) & 0.138 & 0.13 & 0.34 & 2.4 \\
\hline
\end{tabular}
\end{table}

In conclusion, a random matrix model to describe the coupling 
of $m$-fold symmetry 
is constructed. The particular threefold case is used to analyse 
data on eigenfrequencies of elastomechanical vibration of a 
anisotropic quartz block. By properly taking into account the ML 
effect we have shown that the quartz block could very well be
described by 3 uncoupled GOE's , which are gradually coupled by the 
breaking of the three-fold symmetry
(through the gradual removal of octants of increasing sizes), till a
1GOE situation is attained. This, therefore, indicates that the
unperturbed quartz block may posses another symmetry, besides the flip one.
We have also
verified that if a 2GOE description is used, namely, $m=2$ , then an
account of the large-$L$ behaviour of $\Delta_{3}$ can also be obtained if a
much larger number of levels were missing in the sample. In our
particular case of Fig. 2b, we obtained $g = 0.18$. This is 3 times larger
than the ML needed in the 3GOE description. We consider the large value
of $g$ needed in the 2GOE description, much too large
to conform to the reported data in Ref \cite{Elleg}.
A preliminary version of the formal aspect of this work has appeared in
\cite{last}.

\newpage

\begin{figure}[h]
\includegraphics[width=\textwidth,angle=270]{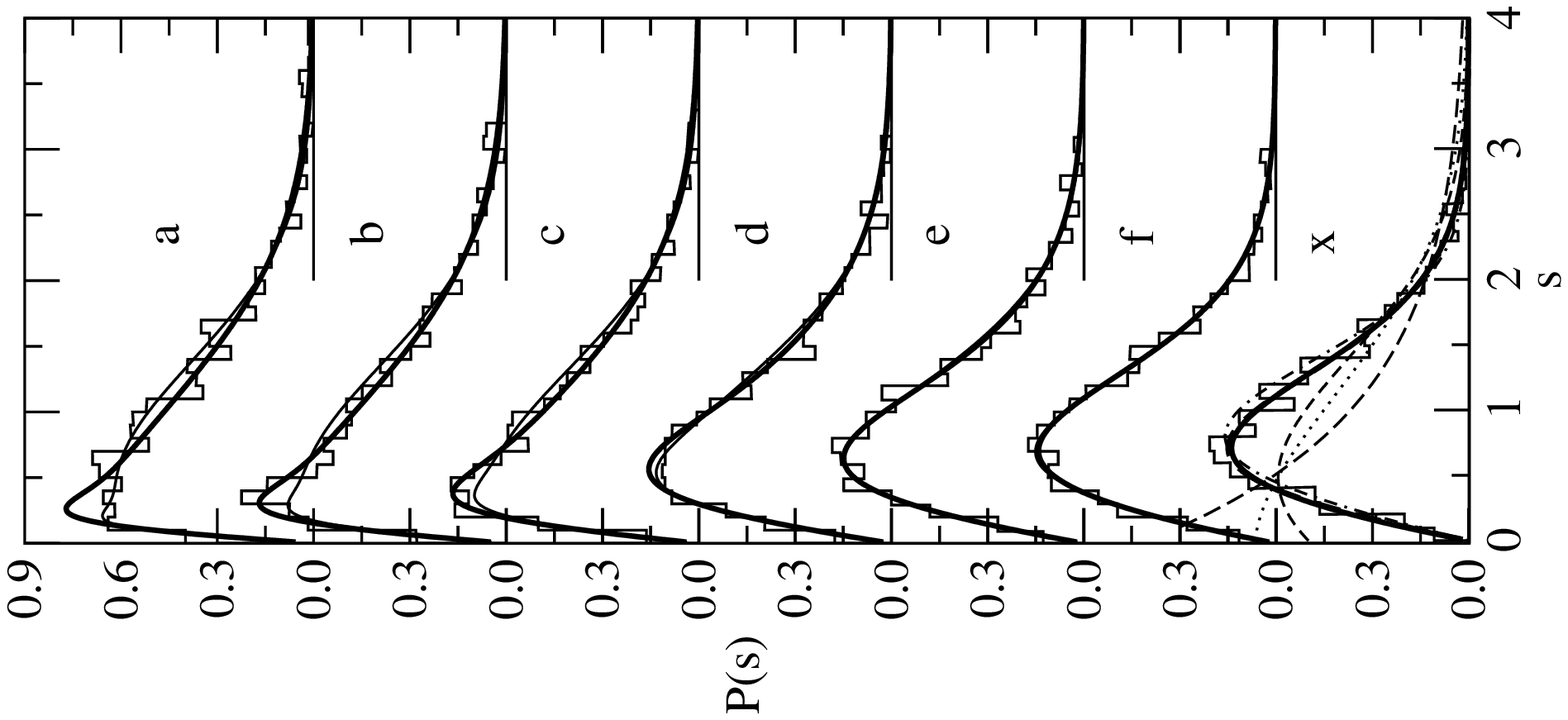}
\caption{
Nearest Neighbour Distributions. Histograms show data (a)-(x) from
Ref. \cite{Elleg}. Thin and thick solid lines show fits to the data
carried out using Eq. (\ref{verde}) with m=2 and m=3 respectively. In
graph (x) the long-dashed line is the Poisson distribution, the dot-dashed
line is the Wigner distribution and the dashed and dotted lines are
the respective distributions for superpositions of 2 and 3 uncoupled GOEs.
See Table I for the values of $\Lambda$ obtained from the fits and the
text for details.
} \label{eterno}
\end{figure}
\begin{figure}
\includegraphics[width=\textwidth,angle=270]{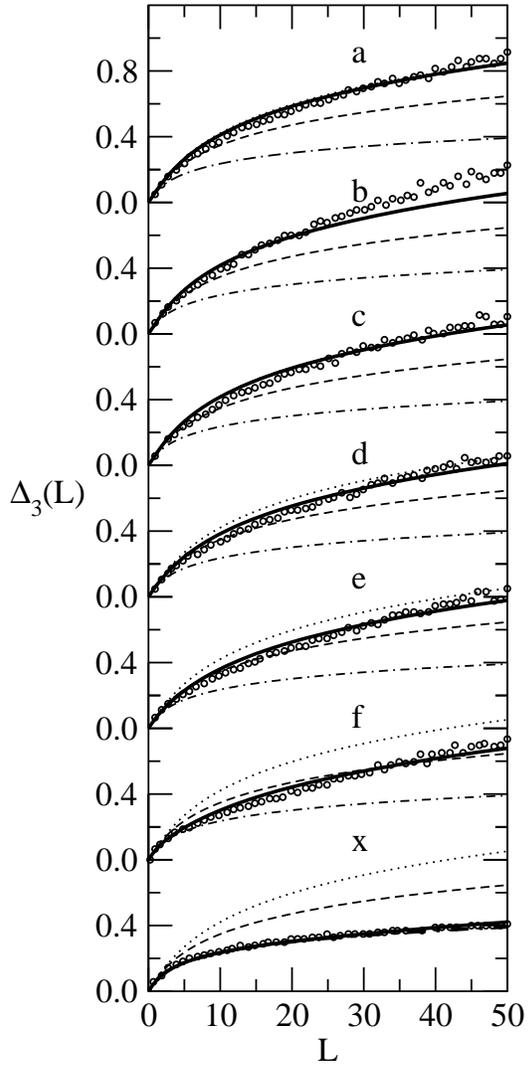}
\caption{
Spectral Rigidities. Circles show data (a)-(x) from
Ref. \cite{Elleg}. Thick solid lines show fits to the data
carried out using Eq. (\ref{verde}) with m=3. The dot-dashed
line is the GOE spectral rigidity and the dashed and dotted lines are
the respective rigidities for superpositions of 2 and 3 uncoupled GOEs.
See Table I for the values of $\Lambda$ obtained from the fits.
} \label{presente}
\end{figure}
\begin{figure}
\includegraphics[width=\textwidth,angle=360]{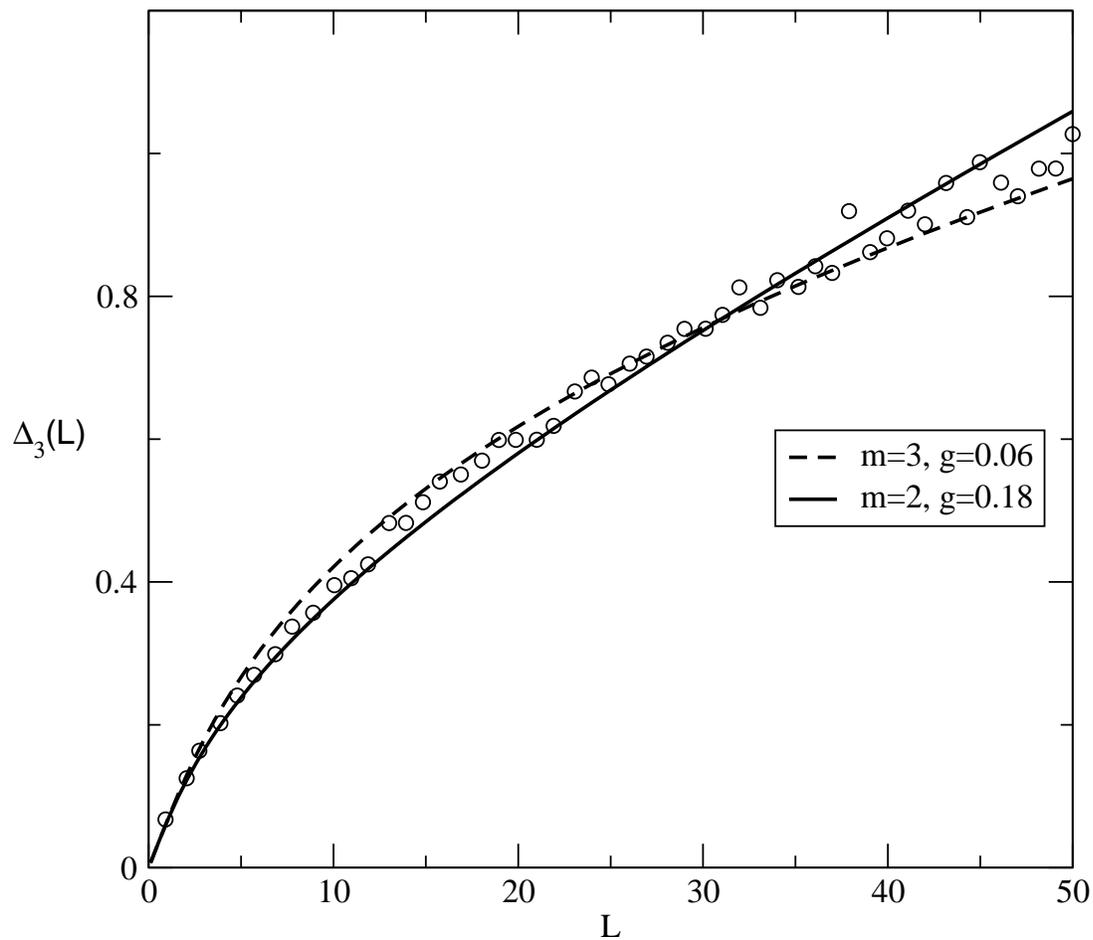}
\caption{The Missing Level Effect: The full curve is the 2GOE $\Delta_3$ for case 2b with 18\% of the levels
missing, while the dashed curve corresponds to 3GOE with only 6\% of the levels missing. The data points (open 
circles) represent fig. 2b. See text for details. 
} \label{presente3}
\end{figure}

\end{document}